\documentclass[aps,prd,twocolumn,amsfonts,amsmath,amssymb,final,footinbib,longbibliography,10pt]{revtex4-2}
\usepackage[dvipsnames]{xcolor}
\usepackage[colorlinks=true,allcolors=Blue]{hyperref}
\usepackage{upgreek,graphicx,manfnt,pifont,colonequals,bbm,stmaryrd}
\usepackage[shortlabels]{enumitem}

\DeclareRobustCommand{\msf}[1]{%
  \ifcat\noexpand#1\relax\msfgreek{#1}\else\mathsf{#1}\fi
}
\makeatletter
\newcommand{\msfgreek}[1]{\csname s\expandafter\@gobble\string#1\endcsname}
\makeatother

\DeclareSymbolFont{sfletters}{OML}{cmbrm}{m}{it}
\DeclareMathSymbol{\seta}{\mathord}{sfletters}{"11}

\newcommand\noi{\noindent}

\newcommand\hf{\frac{1}{2}}

\newcommand\bC{{\mathbb C}}
\newcommand\bR{{\mathbb R}}

\newcommand\bV{{\mathbb V}}
\newcommand\bZ{{\mathbb Z}}
\newcommand\BB{{\mathcal B}}
\newcommand\CC{{\mathcal C}}
\newcommand\DD{{\mathcal D}}

\newcommand\MM{{\mathcal M}}
\newcommand\NN{{\mathcal N}}
\newcommand\OO{{\mathcal O}}

\renewcommand\SS{{\mathcal S}}

\newcommand\UU{{\mathcal U}}

\newcommand\WW{{\mathcal W}}

\newcommand\af{\mathfrak{a}}
\newcommand\cf{\mathfrak{c}}

\newcommand\gf{\mathfrak{g}}
\newcommand\glf{\mathfrak{gl}}
\newcommand\slf{\mathfrak{sl}}
\newcommand\suf{\mathfrak{su}}

\newcommand\uspf{\mathfrak{usp}}
\newcommand\rmUSp{\mathrm{USp}}
\newcommand\rmSU{\mathrm{SU}}
\newcommand\rmU{\mathrm{U}}

\newcommand\rmSO{\mathrm{SO}}

\newcommand\restr[2]{{\left.\kern-\nulldelimiterspace#1\vphantom{\big|}\right|_{#2}}}
\newcommand\ie{\emph{i.e.}}
\newcommand\cff{\emph{c.f.}}
\newcommand\eg{\emph{e.g.}}

\newcommand\kZ{\ell}

\newcommand\tagrel[2]%
  {\stackrel{\scriptscriptstyle(\mkern-1.5mu#1\mkern-1.5mu)}{#2}}

\newcommand{\overbar}[1]{\mkern 2mu\overline{\mkern-2mu#1\mkern-2mu}\mkern 2mu}

\begin{document}
\title{A geometric free field realisation for the genus-two class \texorpdfstring{$\SS$}{S} theory of type \texorpdfstring{$\af_1$}{a1}}

\author{Christopher Beem}
\affiliation{Mathematical Institute, University of Oxford, Woodstock Road, Oxford, OX2 6GG, United Kingdom}
\affiliation{St. John's College, University of Oxford, St. Giles', Oxford, OX1 3JP, United Kingdom}

\author{Carlo Meneghelli}
\affiliation{Dipartimento SMFI, Universit\`a di Parma, Viale G.P. Usberti 7/A, 43100, Parma, Italy}
\affiliation{INFN Gruppo Collegato di Parma}

\date{\today}

\begin{abstract}
\noi We present a free field realisation for the vertex operator algebra associated to the genus-two, class $\SS$ superconformal field theory of type $\af_1$. The free field realisation is in the style of recent work by the authors, and is formulated in terms of a one-dimensional isotropic lattice vertex algebra along with two pairs of symplectic fermions. Our realisation makes manifest an enhanced $\rmUSp(4)$ outer automorphism group of the VOA that is inherited from the symplectic fermion system. This extends an $\rmSU(2)$ outer automorphism that has been observed in recent work of Kiyoshige and Nishinaka and significantly simplifies the structure of the algebra. Along the way, we also produce a realisation of the generic subregular Drinfel'd--Sokolov $\WW$ algebra of type $\cf_2$ in terms of the generic principle $\WW$ algebra of type $\cf_2$ and a one-dimensional isotropic lattice vertex algebra.
\end{abstract}

\maketitle


To any four-dimensional $\NN=2$ superconformal field theory (SCFT) one may canonically associate a vertex operator (super)algebra (VOA) by restriction to the cohomology of a particular conformal supercharge \cite{Beem:2013sza}, or equivalently, by introducing a certain $\Omega$ background that deforms the holomorphic-topological twist of the theory \cite{Oh:2019bgz,Jeong:2019pzg}. These VOAs have proven to be both revealing windows into the physics of the underlying four-dimensional SCFTs and interesting mathematical-physical objects in their own right.

A notable feature of the VOAs that arise via this correspondence is their relative simplicity. Many four-dimensional theories whose conformal phases are otherwise challenging to study (due to being strongly coupled and not admitting transparent Lagrangian descriptions) turn out to have associated VOAs of the simplest types, such as affine current VOAs and rational Virasoro and $\WW$-algebras. More generally, an important aspect of the associated VOAs is their close relationship to the geometry of the Higgs branch of vacua. This has been formalised in the Higgs Branch Conjecture of \cite{Beem:2017ooy}, which in particular implies that these VOAs are all quasi-Lisse \cite{Arakawa:2016hkg}.

Starting with \cite{Beem:2019tfp} (see also the earlier work \cite{Bonetti:2018fqz} and further developments in \cite{Beem:2019snk}), it has emerged that there often exist parsimonious descriptions of the associated VOAs through geometrically motivated free field realisations. Though the physical principles underlying these free field realisations have yet to be completely elucidated, the intuitive picture that has emerged is that the physics of the Higgs branch (as encoded in the Higgs branch as a holomorphic symplectic variety and the residual degrees of freedom present in generic Higgs branch vacua) largely determines the form of a free field construction of geometrically meaningful operators (\eg, Higgs branch chiral ring operators), and the remaining strong generators make themselves known upon consideration of the singular terms in the operator product expansions (OPE) among those geometric operators \footnote{We recall that a strong generator of a VOA is, by definition, any (quasi-primary) operator that cannot be written as the normally ordered product of any other collection of operators and their derivatives.}. In practice, the methodology of this approach remains a mixture of art and science.

In this note, we present an interesting new instance of such a free field realisation, this time for the VOA associated to the class $\SS$ theory of type $\af_1$ for a genus-two surface with no punctures. This is a Lagrangian gauge theory (with gauge group $\rmSU(2)\times\rmSU(2)\times\rmSU(2)$), and the formulation of the VOA as a BRST quotient received some consideration in \cite{Beem:2013sza}. A presentation as a strongly finitely generated vertex algebra has recently been derived in \cite{Kiyoshige:2020uqz} using a combination of BRST cohomology and VOA bootstrap methods. The result is fairly complicated, with seventeen strong generators organised into eleven irreducible representations of a novel $\rmSU(2)$ outer automorphism group.

By contrast, the free field realisation we find is quite simple. It utilises an isotropic lattice vertex subalgebra of a lattice VOA of signature $(1,1)$, along with two pairs of symplectic fermions. This choice of ingredients follows, according to the general calculus of our free field strategy, from the fact that the Higgs branch of the genus two theory is the $D_3$ Kleinian singularity (so it has quaternionic dimension one), and the residual degrees of freedom on the Higgs branch are a pair of Abelian vector multiplets (each giving rise to a symplectic fermion pair). The overall shape of our construction shares a number of qualitative features with the examples studied in our previous work, but there are important differences in the details that we hope point towards broader generalisations of the method.

The organisation of the rest of the paper is as follows. In Section \ref{sec:review} we introduce and review salient aspects of the genus-two SCFT in question. In Section \ref{sec:FFR} we motivate and present the free field realisation of the genus-two vertex algebra. Our realisation makes manifest a (surprisingly large) $\rmUSp(4)$ outer automorphism symmetry that extends the previously identified $\rmSU(2)$ and is inherited directly from the symplectic fermions. We use this to give a more compact presentation of the VOA in terms of the OPEs of strong generators than that of \cite{Kiyoshige:2020uqz}. In Section \ref{sec:observations} we elaborate on a number of technical aspects of our construction. These include the existence of a subVOA of type $\WW[1,2,2,2]$ that is realised in terms of the lattice bosons and a $\WW[2,4]$ subalgebra of the symplectic fermions. We identify this with a special case of the subregular Drinfel'd--Sokolov $\WW$-algebra of type $\cf_2$, and we find an extension of our construction that provides a realisation of the same Drinfel'd--Sokolov $\WW$-algebra at generic level in terms of the same lattice bosons as well as a generic $\WW[2,4]$ algebra. This construction is in many ways analogous to recent work on the subregular Drinfel'd--Sokolov $\WW$-algebra of type $\af_2$ \cite{Adamovic:2020ktz}. As an application of our results, in Section \ref{sec:Higgs_branch_model} we describe the canonical $R$ filtration on our free field vertex algebra and use it to predict a simple realisation of the Higgs branch and Hall-Littlewood chiral rings of the theory in the style of \cite{Beem:2019tfp}. We make additional observations and highlight open questions in Section \ref{sec:discussion}.


\begin{figure}[t]
\includegraphics[width=0.315\textwidth]{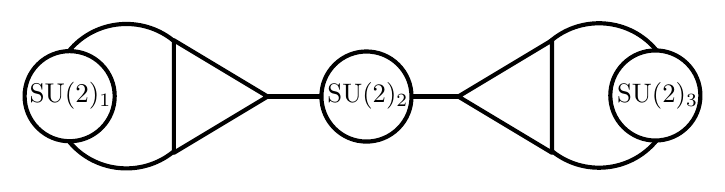}
\includegraphics[width=0.2\textwidth]{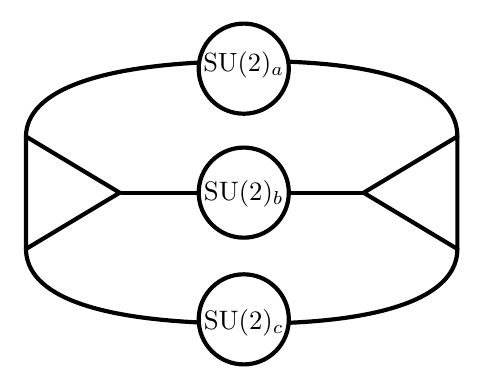}
\caption{\label{fig:quivers}Generalised quiver diagrams for the two duality frames of the genus-two SCFT of type $\af_1$. Triangles (depicting trivalent vertices) correspond to half-hypermultiplets in the tri-fundamental representation of $\rmSU(2)^3$.}
\end{figure}

\section{\label{sec:review}Review of the genus-two theory}

The theory of interest in this work is the genus-two theory of class $\SS$ in type $\af_1$ \cite{Gaiotto:2009we,Gaiotto:2009hg}. This theory admits two $S$-dual Lagrangian realisations that are represented by generalised quiver diagrams with two trivalent vertices corresponding to half-hypermultiplets in the tri-fundamental representation of $\rmSU(2)\times\rmSU(2)\times\rmSU(2)$, see Fig. \ref{fig:quivers}. These give rise to two (inequivalent) realisations of the associated VOA as an $\suf(2)\times\suf(2)\times\suf(2)$ BRST quotient of eight copies of the $\beta\gamma$ (symplectic boson) VOA.

This Lagrangian perspective was touched upon briefly in \cite{Beem:2013sza} and has more recently been studied in detail in work of Kiyoshige and Nishinaka \cite{Kiyoshige:2020uqz}, which led to a proposal for the associated VOA presented in terms of the OPEs of a set of strong generators. For our purposes, the important features of this theory are not connected directly to its Lagrangian realisation, and instead pertain to the structure of the Higgs branch of moduli space, the corresponding low energy moduli space dynamics, and select data that can be extracted from anomalies and the superconformal index. We review and collect all of the necessary information in the remainder of this section.

\vspace{-10pt}

\subsection{\label{subsec:higgs_branch}Higgs branch}

\vspace{-2pt}

The Higgs branch of the vacuum moduli space for this theory has previously been determined to be the Du Val/Kleinian singularity of type $D_3\cong A_3$ \cite{Hanany:2010qu}. This is a quaternionic-dimension-one hyperk\"ahler cone that is represented as a hypersurface singularity in $\bC^3$ according to,
\begin{equation}
\MM_{\rm H} = \left\{\;({\sf x}, {\sf y}, {\sf z})\in\bC^3 ~~\middle|~~ {\sf x}{\sf y} - {\sf z}^4=0\;\right\}~.
\end{equation}
The coordinate ring of this algebraic variety is identified with the Higgs branch chiral ring of the genus-two SCFT. The holomorphic symplectic form on $\MM_{\rm H}$ induces a Poisson structure on the coordinate ring, which (with a particular normalisation of our choice) is defined by the following brackets for the basic coordinate functions,
\begin{equation}\label{eq:Higgs_Poisson}
\{{\sf x},{\sf y}\} = -8\,{\sf z}^3~,~~ \{{\sf y},{\sf z}\} = 2\,{\sf y}~,~~ \{{\sf z},{\sf x}\} = 2\,{\sf x}~.
\end{equation}

This theory is exceptional amongst the higher-genus class $\SS$ theories without punctures in that it has a(n Abelian) flavour symmetry $\rmU(1)_F$ \footnote{At higher genus in type $\af_1$, the Higgs branch of the genus $g$ theory is the $D_{g+1}$ Kleinian singularity. It is only due to the exceptional isomorphism $D_3\cong A_3$ that the genus-two theory manages to enjoy a continuous flavour symmetry.}. The holomorphic moment map for the action of this symmetry on the Higgs branch is the coordinate function ${\sf z}$, which in the Higgs chiral ring corresponds to the superconformal primary in a $\hat{\BB}_1$ conserved current multiplet. In the normalisations of \eqref{eq:Higgs_Poisson} we therefore assign ${\sf x}$ and ${\sf y}$ to have $\rmU(1)_F$ charges $+2$ and $-2$, respectively.

Of particular interest for the development of our free field construction is a Zariski open subset of the Higgs branch, $\UU_{\sf x}\subset\MM_{\rm H}$, that we define as
\begin{equation}
\UU_{\sf x} = \left\{\;{\sf q}\in\MM_{\rm H} ~~\middle|~~{\sf x}({\sf q})\neq0\;\right\}~.
\end{equation}
The coordinate ring of $\UU_{\sf x}$ is the localisation of $\bC[\MM_{\rm H}]$ at ${\sf x}$, $\bC[\UU_{\sf x}]\cong\bC[{\sf x},{\sf x}^{-1},{\sf z}]$ (in this patch, one realises ${\sf y}$ as a composite according to ${\sf y}={\sf z}^4{\sf x}^{-1}$). We then have the identification $\UU_{\sf x}\cong\bC^\ast\times\bC$, which can be upgraded to an isomorphism of Poisson varieties,
\begin{equation}
\UU_{\sf x}\cong T^\ast\bC^\ast~.
\end{equation}
If we let ${\sf p}$ be the coordinate for the cotangent fibre obeying the canonical Poisson bracket $\{{\sf p},{\sf x}\}=1$, then we make the identification ${\sf z}= 2{\sf p}$. The Poisson brackets for ${\sf y}$ presented in \eqref{eq:Higgs_Poisson} then follow directly from those of ${\sf x}$ and ${\sf z}$.

\vspace{-10pt}

\subsection{\label{subsec:residual_gauge}Residual \texorpdfstring{$\rmU(1)$'s}{U(1)'s} and an enhanced Higgs branch}

\vspace{-2pt}

The $\rmSU(2)\times\rmSU(2)\times\rmSU(2)$ gauge symmetry is not completely broken on the Higgs branch. That this is so can be seen by simple dimension counting, since the representation space of the hypermultiplets has quaternionic dimension eight and the hyperk\"ahler quotient should naively reduce this by nine. The presence of a one-quaternionic-dimensional Higgs branch means that there must be a two-dimensional unbroken gauge group in generic Higgs branch vacua, namely $\rmU(1)\times\rmU(1)$. The same conclusion follows quickly when considering the realisation of the theory in terms of a pair of parallel ${\rm M}5$ branes wrapping a genus-two curve, where one may go onto the Higgs branch by separating the two branes in a transverse $\bR^3$. The residual $\rmU(1)$ gauge fields then occur as reductions of the self-dual two-form in the Abelian $(2,0)$ theory.

The presence of this residual gauge symmetry everywhere on the Higgs branch means that the entirety of $\MM_{\rm H}$ is, in fact, smoothly embedded into a larger mixed branch, where the scalars in the vector multiplets associated to the unbroken gauge symmetry acquire non-vanishing expectation values \footnote{In \cite{Hanany:2010qu}, the monicker \emph{Kibble branch} was introduced to indicate this state of affairs. We will use a different terminology.}. We will refer to this larger mixed branch as an \emph{enhanced Higgs branch}, in analogy with the enhanced Coulomb branch terminology of \cite{Argyres:2016xmc}. Due to the local factorisation of the moduli space into Higgs and Coulomb branch directions, along with the fact that $\rmU(1)_r$ is preserved on the Higgs branch and gives a contracting $\bC^\ast$ action on the Coulomb branch, the enhanced Higgs branch has the form of a two-complex-dimensional vector bundle fibred with flat connection over the regular locus of the Higgs branch, $\MM^{\rm reg}_{\rm H}\subset\MM_{\rm H}$ (here the regular locus is just the complement of the conformal point at the tip of the cone).

We can say more about the global structure of the enhanced Higgs branch on fairly general grounds. Since the Coulomb fibration is locally trivial, its global structure will ultimately be determined by a two-dimensional unitary representation (the monodromy representation) of the local fundamental group of the Higgs branch, which can be inferred from the realisation of the Higgs branch as an orbifold: $\pi^1(\MM^{\rm reg}_{\rm H})\cong\bZ_4$. Additionally, the two $\rmU(1)$ factors in the residual gauge group will come with different gauge couplings at generic points of the conformal manifold (as they must arise from different subgroups of the maximal torus of the microscopic gauge group $\rmSU(2)\times\rmSU(2)\times\rmSU(2)$, each simple factor of which has a separately marginal gauge coupling). The fibre bundle describing the enhanced Higgs branch will consequently be decomposable as the direct sum of two line bundles, with the corresponding representation of $\bZ_4$ being analogously decomposable as the direct sum of two one-dimensional representations. (In fact, it must be two copies of the same one-dimensional representation for symmetry reasons, though we will not belabour this point here.)

As the $\rmU(1)$ vector multiplets on the enhanced Higgs branch originate in the gauge group of the microscopic theory, the nontrivial monodromy action on the Coulomb fibres factors through the Weyl group $W(\rmSU(2)^3)\cong\bZ_2^3$. Consequently the generator of $\bZ_4$ must act either trivially or by negation on the Coulomb fibres. Which of these two possibilities is realised could in principle be determined by a more careful analysis of the equations for the full moduli space of vacua. Instead, we will pause our analysis here and will see from the free field realisation later that (subject to our overall scheme for constructing the realisation) the only reasonable option is for it to act by negation. At the level of the patch $\UU_{\sf x}$, this means that the enhanced Higgs branch restricts to a two-dimensional Coulomb branch fibration whose connection has a holonomy of $-1$ when transported around the origin in $\bC^\ast$.

\begin{table}[t]
\begin{ruledtabular}
\begin{tabular}{ccccc}
~~~Multiplet~~~                     		& Name  		  &~~~$h$~~~&~~$\rmU(1)_r$~~&~~$\rmU(1)_F$~~\\[2pt]
\hline
$\hat{\BB}_1$                       		& $Z$   							& $1$   & $0$	 & $ 0$	\\[2pt]
$\hat{\BB}_2$                       		& $X$   							& $2$   & $0$	 & $+2$	\\[2pt]
$\hat{\BB}_2$                       		& $Y$   							& $2$   & $0$	 & $-2$	\\[2pt]
\hline
$2\times\DD_{1\,(0,0)}$             		& $b^+_{~\;1}\;,\;b^+_{~\;2}$ 		& $2$  	& $+\hf$ & $+1$	\\[2pt]
$2\times\overbar{\DD}_{1\,(0,0)}$      		& $\bar{b}^{+1}\;,\;\bar{b}^{+2}$ 	& $2$  	& $-\hf$ & $+1$	\\[2pt]
$2\times\DD_{1\,(0,0)}$             		& $b^-_{~\;1}\;,\;b^-_{~\;2}$		& $2$  	& $+\hf$ & $-1$ \\[2pt]
$2\times\overbar{\DD}_{1\,(0,0)}$      		& $\bar{b}^{-1}\;,\;\bar{b}^{-2}$	& $2$   & $-\hf$ & $-1$ \\[2pt]
$\DD_{\frac32 \,\left(0,\hf\right)}$        & $\omega$             				& $3$   & $+1$   & $ 0$ \\
$\overbar{\DD}_{\frac32\,\left(\hf,0\right)}$ & $\overbar{\omega}$ 				& $3$  	& $-1$	 & $ 0$	\\[4pt]
\hline
$\hat{\CC}_{0(0,0)}$                		& $T$ 		        				& $2$ 	& $0$	 & $ 0$	\\[2pt]
\hline
$3\times\hat{\CC}_{1(0,0)}$         		& $P\;,\;Q\;,\;R$ 	        		& $3$ 	& $0$	 & $ 0$	\\
\end{tabular}
\end{ruledtabular}
\caption{\label{tab:genus2gens}Chiral algebra generators for the genus two theory. The first two columns list the four-dimensional multiplet and name for each generator. The remaining columns list their two-dimensional quantum numbers.}
\end{table}

\vspace{-10pt}

\subsection{\label{subsec:voa_generators}VOA generators}

\vspace{-2pt}

A list of strong generators of the genus two vertex algebra with two-dimensional conformal weight $h\leqslant3$ was determined in \cite{Beem:2013sza} on the basis of an analysis of the superconformal index. This list was subsequently shown to constitute a complete set of strong generators in the analysis of \cite{Kiyoshige:2020uqz}. We reproduce this list in Table \ref{tab:genus2gens} along with our naming conventions for the generators added.

The $\hat\BB$-type operators $X$, $Y$, and $Z$ descend from the Higgs branch chiral ring operators in four dimensions that correspond respectively to the coordinate functions ${\sf x}$, ${\sf y}$, and ${\sf z}$ on $\MM_{\rm H}$. In particular, $Z$ is related to the moment map on the Higgs branch and so generates a $\widehat{\glf(1)}$ affine current subalgebra. The level of this current algebra (\ie, the coefficient of the quadratic pole in its self-OPE in the normalisation implied by the charge assignments in Table \ref{tab:genus2gens}) can be read off from the field content in the Lagrangian description of the theory and is given by $\kZ=-2$ \footnote{We reserve the character $k$ for the levels of non-Abelian affine current algebras as appear later in this paper.}.

The operators $b^+_{~\;a}$ and $b^{-}_{~\;a}$ ($a=1,2$) and $\omega$ are Hall-Littlewood chiral ring generators while $\bar{b}^{+a}$ and $\bar{b}^{-a}$, and $\overbar{\omega}$ are their Hall-Littlewood anti-chiral ring counterparts. $T$ is the VOA stress tensor, which generates a Virasoro subalgebra. Like with the affine current level, the Virasoro central charge is determined by the Lagrangian field content of the theory (or alternatively, by the four-dimensional $c$-type Weyl anomaly coefficient), and is given by $c=-26$.

The additional generators $P$, $Q$, and $R$ are more exotic since they are non-stress tensor, $\hat{\CC}$-type operators. The existence was deduced in \cite{Beem:2013sza} from an analysis of the superconformal index at relatively low order. In the free field realisation described in the next section, these will be observed to arise automatically in the $b\times\bar{b}$ OPE.

\vspace{-10pt}

\section{\label{sec:FFR}The free field realisation}

\vspace{-2pt}

In this section we derive and present a free field realisation for the genus-two VOA. While an explicit form for the VOA in terms of strong generators has been given in \cite{Kiyoshige:2020uqz}, we will operate behind a veil of ignorance regarding results so as to help illustrate how the EFT-motivated free field Ansatz constrains the form of the free field realisation, ultimately leading to a quick and efficient (re-)derivation of the full OPEs. For reference, note that the complete set of OPEs (in our conventions, and in terms of strong generators as redefined in Equations \eqref{eq:bchiredefinition}, \eqref{eq:upsilon_redefinition}, and \eqref{eq:upsilon_redefinition_two}) are given by canonical $T$ and $Z$ OPEs encoding dimensions and $\rmU(1)_F$ charges, Equations \eqref{eq:chi_covariant_OPEs}, \eqref{eq:chi_big_OPE}, and \eqref{eq:covariant_OPEs}, and the $\bZ_2$ transformations of those with respect to the automorphism given in \eqref{eq:Z2_automorphism}.

We adopt the philosophy developed in our previous work \cite{Beem:2019tfp} and introduce ingredients in accordance with the Higgs branch physics of the theory. To this end, in correspondence with the $\UU_{\sf x}$ patch of $\MM_{\rm H}$ and the two residual vector multiplets we introduce the free field VOA,
\begin{equation}\label{freefieldIngredients}
\bV_{\rm free}=\mathrm{SF}(2)\otimes \Pi~.
\end{equation}
The free field ingredients on the right hand side are defined as follows. The factor $\mathrm{SF}(2)$ denotes the VOA of two pairs of symplectic fermions (denoted here by $\eta$),
\begin{equation}\label{SimpFERM}
\eta_A(z)\,\eta_B(w)\sim\,\frac{\Omega_{AB}}{(z-w)^2}~,\quad A,B=1,2,\,\dots,4~,
\end{equation}
where $\Omega_{AB}$ is a non-degenerate skew-symmetric matrix (the symplectic form). Whenever relevant, we will take $\Omega$ to have a canonical form with $\Omega_{13}=\Omega_{24}=1$ and other entries not related by skew-symmetry vanishing. The stress tensor for the symplectic fermions is given by
\begin{equation}
\label{stresstensorFermions}
T_{\eta}\colonequals-\frac{1}{2}\Omega^{AB}\eta_A\eta_{B}~,
\end{equation}
and has central charge $c_{\eta} = -4$. The VOA \eqref{SimpFERM} enjoys a $\rmUSp(4)$ group of outer automorphisms (which accounts for the name of the algebra) under which the generators transform in the ${\bf 4}$.

The factor $\Pi$ denotes an ``isotropic lattice VOA'' (see \cite{Berman2001RepresentationsOA,Adamovic:simple_affine}). This is realised in terms of two chiral bosons $\delta(z),\varphi(z)$ with nonvanishing OPEs
\begin{equation}\begin{split}\label{deltaphiOPE}
\delta(z_1)\delta(z_2)&\sim \langle \delta,\delta\rangle\,\log z_{12}~,\\
\varphi(z_1)\varphi(z_2)&\sim \langle \varphi,\varphi\rangle\,\log z_{12}~,
\end{split}
\end{equation}
where $z_{12}=z_1-z_2$ and we take $\langle \delta,\delta\rangle = -\langle \varphi,\varphi\rangle$, so that $\delta+\varphi$ is null. The vertex algebra $\Pi$ is the subalgebra of the rank $(1,1)$ lattice VOA associated to these bosons where the lattice momentum is restricted to the null direction $\delta+\varphi$ \footnote{The choice of half-integer values for the lattice momentum is a convention that, at this stage, does not sacrifice any generality; a change of convention could be absorbed by a rescaling of the normalisations of the lattice bosons. Our choice serves to facilitate a simple identification of the operator $X$ in \eqref{positeiveGENgenus2nopunct}.},
\begin{equation}\label{PilattexVOA}
\Pi\colonequals \bigoplus_{n=-\infty}^{\infty}\,\left(V_{\partial\varphi}\otimes V_{\partial\delta}\right)\,e^{\frac{n(\delta+\varphi)}{2}}~,
\end{equation}
where $V_j,~j\in\{\partial\varphi,\partial\delta\}$ is the $\widehat{\glf(1)}$ affine current VOA associated with the current $j$ \footnote{$\Pi$ can be thought of as a kind of $\beta\gamma$ vertex algebra with (say) $\gamma$ made invertible, and indeed this is a very useful formulation for performing computations. Alternatively, with appropriate normalisations, this can interpreted as the algebra of chiral differential operators on the multiplicative group $\bC^\ast$. We thank Sujay Nair for pointing out the latter identification.}.

We can distill the requirements that were imposed on similar free field constructions in \cite{Beem:2019tfp} and apply them to the present case. We arrive at the following operational rules that we will impose on the putative free field realisation of the genus-two VOA in $\Pi\otimes{\rm SF}(2)$:
\begin{enumerate}[(i)]
\item The affine current $Z$ should be identified with $\partial\varphi$ up to an overall rescaling. $\rmU(1)_F$ charge is then related to the lattice momentum appearing in the exponentials in $\Pi$ \footnote{As long as we take $\varphi$ to be the chiral boson with negative norm, there is no loss of generality here compared to taking a general (non-null) linear combination of $\partial\varphi$ and $\partial\delta$. This is because the lattice vertex algebra enjoys the usual $\rmSO(1,1)$ automorphism by hyperbolic rotations between the bosons. Our conventions regarding this freedom differ from those of \cite{Adamovic:2020ktz}, where in a similar situation the normalisations of the bosons $\delta$ and $\varphi$ were fixed to conventional values but the affine current had to be taken to be more complicated linear combination so as to realise the correct affine current level.}.
\item The four-dimensional $\rmU(1)_r$ symmetry is identified with the subgroup of the $\rmUSp(4)$ symmetry of the symplectic fermions under which $\eta_1$ and $\eta_2$ have charge $+1$ and $\eta_3$ and $\eta_4$ have charge $-1$ \footnote{Said differently, we are taking $\eta_{1}$ and $\eta_2$ to correspond to four-dimensional gauginos of the type $\lambda^1_{+}$, while $\eta_3$ and $\eta_4$ correspond to gauginos of the type $\tilde\lambda^1_{\dot{+}}$.}. This is an exact symmetry of the VOA.
\item The VOA stress tensor is the sum of canonical stress tensors for the symplectic fermions and for $\varphi$, plus a stress tensor for $\delta$ that may include some non-zero background charge.
\item The affine current $\partial\varphi$ and its derivatives do not appear in the expression for VOA generators other than $Z$ and $T$ as prescribed above \footnote{This enforces the requirement that the other strong generators be primaries with respect to the affine current.}.
\item The Higgs branch generator $X$ is realised with the operator $e^{\delta+\varphi}$. This plays the role of an affine version of the invertible coordinate ${\sf x}$ in $\UU_{\sf x}$. Requiring the correct conformal weight for this operator fixes the value of the background charge for $\delta$ in the stress tensor.
\item The symplectic fermions are understood as being valued in a fermionic vector bundle over $\bC^\ast$ associated to the Coulomb branch fibration of the enhanced Higgs branch described above. VOA operators should be constructed from ``single-valued'' quantities on $\UU_{\sf x}$, taking account of the monodromy of the Coulomb fibres.
\end{enumerate}

The details of our theory serve to immediate specialise some of the above considerations. In order to match $\rmU(1)_F$ charge normalisations with those of Table \ref{tab:genus2gens}, we identify
\begin{equation}
Z = \frac{2}{\langle\varphi,\varphi\rangle}\partial\varphi~.
\end{equation}
Consequently, the level $\kZ$ of the affine current is related to the bosonic normalisation,
\begin{equation}
Z(z_1)Z(z_2)\sim\frac{\kZ}{(z_{12})^2}~,\quad \kZ= \frac{4}{\langle\varphi,\varphi\rangle}~.
\end{equation}
Charge considerations (in conjunction with the general scheme presented above) then fixes the form of the operators with positive $\rmU(1)_F$ charges to take extremely simple forms. In particular, in order to be able to assign the correct $\rmU(1)_F$ charge to the Hall-Littlewood operators $b^+_{~\;a}$ and $\bar{b}^{+a}$, we must let the symplectic fermions to have $\bZ_2$ monodromy (resolving the question of the global structure of the enhanced Higgs branch), which allows (and requires) odd numbers of symplectic fermions to be accompanied by half-integer powers of $e^{\delta+\varphi}$. We then have the following simple realisations of the positive-charge Higgs and Hall-Littlewood generators,
\begin{equation}\label{positeiveGENgenus2nopunct}
X=e^{\delta+\varphi}~,\quad b^+_a=\eta_a\,e^{\frac{\delta+\varphi}{2}}~,\quad \bar{b}^{+a}=\eta^a\,e^{\frac{\delta+\varphi}{2}}~,
\end{equation}
for $a=1,2$, and where we have introduced raised-index symplectic fermions $\eta^A = \eta_B\Omega^{BA}$, so in particular $\eta^1=\eta_3$ and $\eta^2=\eta_4$.

To assign the correct scaling dimensions to these operators, the background charge for $\delta$ is then fixed in terms of the normalisation of $\varphi$, leading to a total stress tensor of the form
\begin{equation}
T = T_{\eta} + \frac{1}{2\langle\varphi,\varphi\rangle}\left((\partial\varphi)^2 - (\partial\delta)^2 + 4\partial^2\delta\right)~.
\end{equation}
The Virasoro central charge for this stress tensor takes the value $c=-2+48/\langle\varphi,\varphi\rangle$. For now we leave the normalisation unfixed, though it is already clear that assigning $\langle\varphi,\varphi\rangle = -2$ will return the correct value for $c$ (and for $\kZ$).

The subset of strong generators with positive $\rmU(1)_F$ charge $\{b^+_{~\;a}, \bar{b}^{+b}, X\}$ then obey the simple OPE relation
\begin{equation}\label{OPEbB}
b^+_a(z_1)\,\bar{b}^{+b}(z_2)\,\tagrel{v.p.}{\sim}\,\frac{\delta_a^bX(z_2)}{z_{12}^2}~,
\end{equation}
with all other OPEs being nonsingular. Here and in the following, we adopt the convention that in our OPEs, \emph{we only display the Virasoro primary operators}. This convention is indicated by the symbol $\tagrel{v.p.}{\sim}$. The full OPE can then be reconstructed from the given data using Virasoro symmetry \footnote{This is particularly straightforward when using the Mathematica package \texttt{OPEconf} of K.~Thielemans \cite{Thielemans:1991uw,Thielemans:1994er}. We have used this and the more general \texttt{OPEdefs} packages to perform many of the calculations reported in this paper.}.

The expressions for the $b^-$ and $\bar{b}^-$ operators turn out to be completely fixed by the requirement that they have the correct $\rmU(1)_r\times\rmU(1)_F$ charges and that the $b^-_{~\;a}\times b^-_{~\;b}$ and $\bar{b}^{-a}\times\bar{b}^{-b}$ OPEs be nonsingular. (This regularity requirement can be seen to follow from charge considerations and the list of strong generators with $h\leqslant3$ given in Table \ref{tab:genus2gens}.) The same requirements fix the normalisation of the bosons to be $\langle\varphi,\varphi\rangle=-2$, so the affine current level and Virasoro central charges that were predicted by four dimensional considerations arise \emph{automatically} at this point within our scheme as a consistency condition.

Direct computation then reveals that the $b^-$ and $\bar{b}^-$ operators obey an OPE relation analogous to \eqref{OPEbB} with the replacement $X\to Y$, \ie,
\begin{equation}\label{OPEbBbar}
b^-_{~\;a}(z_1)\,\bar{b}^{-b}(z_2)\,\tagrel{v.p.}{\sim}\,\frac{\delta_a^b Y(z_2)}{z_{12}^2}~.
\end{equation}
Here we have identified the single nontrivial Virasoro primary operator appearing in this OPE with the Higgs branch generator $Y$. This is the only option consistent with charge assignments given in Table \ref{tab:genus2gens}. In terms of free fields, we have to this point the following realisations for $b^-$, $\bar{b}^-$, and $Y$ operators \footnote{Here and throughout this paper, composite operators should be interpreted in terms of nested conformal normal ordering: $\OO_1\cdots\OO_n = (\OO_1(\OO_2(\cdots(\OO_{n-1}\OO_n))))$.},
\begin{widetext}
\begin{equation}
\begin{split}
b^-_{~\;a}&= \Big(\eta_a\left((\partial\delta)^2- \partial^2\delta-2 T_\eta\right)-2\partial\eta_a (\partial\delta)+ \partial^2\eta_a\Big)e^{-\hf(\delta+\varphi)}~, \\
\bar{b}^{-a}&= \Big(\eta^a\left((\partial\delta)^2- \partial^2\delta-2 T_\eta\right)-2\partial\eta^a (\partial\delta)+ \partial^2\eta^a\Big)e^{-\hf(\delta+\varphi)}~, \\
Y &=\Big((\partial\delta)^4-4(\partial^2\delta)(\partial\delta)^2+(\partial^2\delta)^2+2(\partial^3\delta)(\partial\delta)-\tfrac{1}{3}\partial^4\delta-4T_\eta(\partial\delta)^2+4\partial T_\eta(\partial\delta)+\OO^{(4)}_\eta\Big) e^{-(\delta+\varphi)}~.
\end{split}
\end{equation}
\end{widetext}
The operator $\OO^{(4)}_\eta$ appearing in the expression for $Y$ resides entirely in the ${\rm SF}(2)$ algebra can be expressed as
\begin{equation}
\label{W4fermions}
\begin{split}
\OO^{(4)}_{\eta} &= 44(T_\eta T_\eta) - 14 \partial^2T_\eta - 4 W_\eta~,\\
W_\eta &= 12 T_\eta^2+4\Omega^{AB}(\eta_A\partial^2\eta_B)-3\Omega^{AB}\partial\eta_A\partial\eta_B~.
\end{split}
\end{equation}
We will return in the next section to the conceptual status of the $W_\eta$ operator.

We are now in position to make an important observation, which is that the operators $b^+$ and $\bar{b}^+$ and separately the operators $b^-$ and $\bar{b}^-$ can be assembled into irreducible representations of the full $\rmUSp(4)$ automorphism group of the symplectic fermions. Indeed, let us define
\begin{equation}\label{eq:covariant_fours}
\begin{split}
\chi^+_A(z) &= \eta_Ae^{\frac{\delta+\varphi}{2}}~,\\[4pt]
\chi^-_A(z) &= \Big(\eta_A\left((\partial\delta)^2- \partial^2\delta-2 T_\eta\right)\\
&~\qquad\qquad\quad-2\partial\eta_A (\partial\delta)+ \partial^2\eta_A\Big)e^{-\hf(\delta+\varphi)}~,
\end{split}
\end{equation}
which are related to our previous generators according to
\begin{equation}\label{eq:bchiredefinition}
\chi^\pm_{1,2} \equiv b^{\pm}_{1,2}~,\qquad \chi^{\pm}_{3,4} \equiv \bar{b}^{\pm1,2}~.
\end{equation}
The OPEs \eqref{OPEbB} and \eqref{OPEbBbar} can now be rewritten in the manifestly covariant form,
\begin{equation}\label{eq:chi_covariant_OPEs}
\begin{split}
\chi^+_A(z_1)\,\chi^{+}_B(z_2)\,&\tagrel{v.p.}{\sim}\,\frac{\Omega_{AB}X(z_2)}{z_{12}^2}~,\\
\chi^-_A(z_1)\,\chi^{-}_B(z_2)\,&\tagrel{v.p.}{\sim}\,\frac{\Omega_{AB}Y(z_2)}{z_{12}^2}~.\\
\end{split}
\end{equation}
In turn, the operators $\chi^{\pm}_A$ weakly generate the entire VOA; we recover the rest of the strong generators by examining their singular OPEs (and, in principle, iteratively taking more OPEs until the algebra closes). Notably, this implies that the full VOA also inherits the complete $\rmUSp(4)$ outer automorphism symmetry of the symplectic fermions! In practice, beyond those operators that have already been defined, we need only study the $\chi^+\times\chi^-$ OPEs to realise the full list of strong generators. Directly computing, we find the following singular OPE,
\begin{equation}\label{eq:chi_big_OPE}
\begin{split}
\chi^+_A(z_1)\chi^-_B(z_2)&\tagrel{v.p.}{\sim}\\
\Omega_{AB}&\left(\frac{12}{z_{12}^4}-\frac{6Z}{z_{12}^3}+\frac{Z^2_{(v.p.)}}{z_{12}^2}\right)-\frac{2\Upsilon_{AB}}{z_{12}}~,
\end{split}
\end{equation}
where the new generator $\Upsilon_{AB}$ transforms in the ${\bf 5}$-dimensional, two-index anti-symmetric tensor representation of $\rmUSp(4)$. It is realised in terms of free fields as
\begin{equation}
\Upsilon_{AB} = \tfrac12\left((\partial\delta)\eta_{\llbracket A}\eta_{B\rrbracket} - \partial(\eta_{\llbracket A}\eta_{B\rrbracket})\right)~,
\end{equation}
with double brackets denoting antisymmetrisation and removal of the $\Omega$ trace \footnote{So for a two-index tensor operator $\OO_{AB}$, we have $\OO_{\llbracket AB\rrbracket} \colonequals \OO_{AB}-\OO_{BA}+\frac12 \Omega_{AB}\Omega^{CD}\OO_{CD}$.}.

This five-dimensional multiplet combines all of the remaining strong generators from Table \ref{tab:genus2gens}, namely $\{\omega, P,Q,R, \overbar{\omega}\}$. Indeed, specialising the indices $A,B$ to particular values give the VOA operators corresponding to these different four-dimensional multiplets. The additional $\DD$ and $\overbar{\DD}$ operators are given by
\begin{equation}\label{eq:upsilon_redefinition}
\begin{split}
\omega &= \Upsilon_{12} = \eta_1\eta_2(\partial\delta)-\partial(\eta_1\eta_2)~,\\
\overbar{\omega} &= \Upsilon_{34} = \eta_3\eta_4(\partial\delta)-\partial(\eta_3\eta_4)~,
\end{split}
\end{equation}
while for the trio of $\hat\CC_{1(0,0)}$ operators with $h=3$ we have
\begin{equation}\label{eq:upsilon_redefinition_two}
\begin{split}
P   &= \Upsilon_{14} = \eta_1\eta_4(\partial\delta)-\partial(\eta_1\eta_4)~,\\
Q   &= \Upsilon_{13} = \tfrac{1}{2}\left((\eta_1\eta_3-\eta_2\eta_4)(\partial\delta)-\partial(\eta_1\eta_3-\eta_2\eta_4)\right)~,\\
R   &= \Upsilon_{23} = \eta_2\eta_3(\partial\delta)-\partial(\eta_2\eta_3)~.
\end{split}
\end{equation}

With a full complement of strong generators in place, the defining singular OPEs can now be calculated and they automatically take a manifestly $\rmUSp(4)$ covariant form. To simplify the presentation, we will exploit the existence of an additional descrete symmetry (the origin of which lies with $\mathsf{CPT}$ symmetry, or reflection positivity, in four dimensions) \footnote{In general, four dimensional unitarity implies the existence of an order-four automorphism of the VOA that exchanges operators of opposite $\rmU(1)_r$ charge \cite{Beem:2018duj,unitarity_in_progress}. In the present case, this is combined with the $\bZ_4$ subgroup of $\rmUSp(4)$ that acts by
\begin{equation*}
\{\eta_1,\eta_2,\eta_3,\eta_4\} \to \{\eta_3,\eta_4,-\eta_1,-\eta_2\}~.
\end{equation*}
to give the simpler $\bZ_2$ VOA automorphism \eqref{eq:Z2_automorphism}.}. This manifests for us as a $\bZ_2$ automorphism that exchanges, amongst other things, $\DD$ and $\overbar{\DD}$ operators and reverses $\rmU(1)_F$ charges; more precisely the action on strong generators is as follows,
\begin{equation}\label{eq:Z2_automorphism}
\{X,Y,Z,\chi_A^{\pm},\Upsilon_{BC}\}\to\{Y,X,-Z,\chi_A^{\mp},-\Upsilon_{BC}\}~.
\end{equation}
(In retrospect, given this symmetry one could immediately deduce \eqref{OPEbBbar} from \eqref{OPEbB}.)

We will take advantage of this symmetry by giving a slightly abbreviated list of strong generator OPEs, from which the remaining OPEs can be recovered by acting with the $\bZ_2$ automorphism,
\begin{widetext}
\begin{equation}\label{eq:covariant_OPEs}
\begin{split}
X(z_1)Y(z_2)			&\tagrel{v.p.}{\sim}\frac{72}{z_{12}^4}-\frac{72Z}{z_{12}^3}+\frac{32Z^2_{(v.p.)}}{z_{12}^2}-\frac{8Z^3_{(v.p.)}}{z_{12}}~,\\
\chi^-_A(z_1)X(z_2)	&\tagrel{v.p.}{\sim}\frac{6\chi^+_A}{z_{12}^2}+\frac{4 (Z \chi^+_A)_{(v.p.)}}{z_{12}}~,\\
X(z_1)\Upsilon_{AB}(z_2) &\tagrel{v.p.}{\sim}\frac{-2(\chi^+_{\llbracket A}\chi^+_{B\rrbracket})_{(v.p.)}}{z_{12}}~,\\
\Upsilon_{AB}(z_1)\chi^+_C(z_2) &\tagrel{v.p.}{\sim} \frac{3\Omega^{\phantom{+}}_{C\llbracket A}\chi^-_{B\rrbracket}}{z_{12}^3}+
\frac{\Omega^{\phantom{+}}_{C\llbracket A}(Z\chi^-_{B\rrbracket})_{(v.p.)}}{z_{12}^2}-
\frac{\Omega^{\phantom{+}}_{C\llbracket A}((\partial Z)\chi^-_{B\rrbracket})_{(v.p.)}}{z_{12}}~,\\
\Upsilon_{AB}(z_1)\Upsilon_{CD}(z_2) &\tagrel{v.p.}{\sim} {\rm\bf P}^{\mathbbm 1}_{ABCD}\left(\frac{18}{(z_{12})^6}-\frac{Z^2_{(v.p.)}}{(z_{12})^4}+\frac{(\partial^2ZZ)_{(v.p.)}+\frac{1}{2}(\Omega^{AB}\chi^+_A\chi^-_B)_{(v.p.)}}{(z_{12})^2}\right)+\frac{\frac12(\Omega \chi^+\partial \chi^-)_{ABCD}}{z_{12}}~.
\end{split}
\end{equation}
\end{widetext}
In the above equation, $(\partial^n\OO_1\OO_2)_{(v.p.)}$ denotes the Virasoro primary operator appearing at order $(z_{12})^n$ in the $\OO_1\times\OO_2$ OPE \footnote{This is implemented easily in Mathematica with \texttt{OPEconf} as \texttt{OPEPPole[-n][$\OO_1$,$\OO_2$]}.}. In the final line, ${\rm\bf P}^{\mathbbm 1}_{ABCD}$ is the projection tensor from the ${\bf 5}\otimes{\bf 5}$ of $\rmUSp(4)$ to the singlet,
\begin{equation}
{\rm\bf P}^{\mathbbm 1}_{ABCD} = \Omega_{AC}\Omega_{BD}-\Omega_{AD}\Omega_{BC}-\frac12 \Omega_{AB}\Omega_{CD}~.
\end{equation}
Finally, for a two-index tensor operator $\OO_{AB}$ we have defined
\begin{equation}
\begin{split}
(\Omega\OO)_{ABCD} =\Omega_{AC}&\OO_{(BD)}+\Omega_{BD}\OO_{(AC)}\\
&-\Omega_{AD}\OO_{(BC)}-\Omega_{BD}\OO_{(AC)}~,
\end{split}
\end{equation}
with parentheses denoting symmetrisation (with weight one).

\vspace{-10pt}

\section{\label{sec:observations}Technical observations}

\vspace{-2pt}

There are a number of remarkable features of this VOA and the associated free field realisation. The first, and perhaps most striking, is the observation from the previous section that the full $\rmUSp(4)$ symmetry of the symplectic fermions is inherited by the VOA. This goes beyond the $\rmSU(2)$ symmetry that was observed in \cite{Kiyoshige:2020uqz}, with that symmetry arising as the subgroup of $\rmUSp(4)$ that preserves the spaces of Hall-Littlewood chiral ring operators and Hall-Littlewood anti-chiral ring operators, respectively. (In our conventions, these are block-diagonal $\rmUSp(4)$ matrices with two blocks of size two. Alternatively, this is the centraliser of $\rmU(1)_r\subset\rmUSp(4)$, whose generator is realised by the matrix $\frac12{\rm diag}(+1,+1,-1,-1)$.) This has the surprising consequence of relating VOA operators associated to $\DD$, $\overbar\DD$, and $\hat\CC$ multiplets in four dimensions. Indeed, it is an entertaining consequence of this symmetry that while the genus-two VOA has strong generators beyond the stress tensor and Hall-Littlewood chiral ring operators (which were identified as a somewhat canonical set of generators in \cite{Beem:2013sza}), all of its generators are related to those by outer automorphism symmetry! Additionally, the rather large collection of strong generators of the VOA now organise into just \emph{five} irreducible representations of the full outer automorphism group.

A more technical observation is that the Higgs branch generators and the stress tensor $\{X, Y, Z, T\}$ collectively strongly generate a closed vertex operator subalgebra. Retrospectively, given the list of generators in Table \ref{tab:genus2gens}, the existence of this closed sublagebra is guaranteed \emph{given the existence of enhanced $\rmUSp(4)$ symmetry} (or even the smaller $\rmSU(2)$ symmetry of \cite{Kiyoshige:2020uqz}), since this precludes the appearence of the $\hat\CC_{1(0,0)}$ generators in the $X\times Y$ OPE. This subalgebra is a $\WW$-algebra of type $\WW[1,2,2,2]$, and one may verify by direct computation that it can be identified with (a quotient of) the specialisation of the subregular Drinfel'd--Sokolov $\WW$-algebra for Lie algebra $\cf_2\cong\uspf(4)$ at level $k=-\tfrac{5}{2}$ \footnote{It may be worth noting that in \cite{Beem:2017ooy}, another VOA of type $\WW[1,2,2,2]$ arose in connection with the $(A_1,A_7)$ Argyres-Douglas SCFT. That algebra was identified as the subregular Drinfel'd--Sokolov $\WW$-algebra of type $\af_3$, specialised to the degenerate level $k=-16/5$ where an additional dimension-four strong generator that would generically be present becomes redundant. For the subregular reduction of $\cf_2$, the algebra is generically of type $\WW[1,2,2,2]$.}.

The realisation of this $\WW[1,2,2,2]$ subalgebra utilises the symplectic fermions only through the combinations $T_\eta$ and $W_{\eta}$ introduced in \eqref{stresstensorFermions}, \eqref{W4fermions}. These composite operators are precisely the strong generators of a vertex operator subalgebra $\WW[2,4]\cong {\rm SF}(2)^{\rmUSp(4)}$ \cite{Creutzig:2014xea}. This two-generator VOA is equivalent to the specialisation (and simple quotient) of the principle $\WW$ algebra of type $\cf_2$, $\WW_{-\frac{5}{2}}(\cf_2, f_{\rm prin})$ corresponding to $c=-4$, see Theorem 5.1 in \cite{Kanade:2018qut}. In this sense, ignoring the origins of the operators $T_\eta$ and $W_\eta$ in symplectic fermions, we have a realisation of $\WW_{-5/2}(\cf_2, f_{\rm subreg})$ in terms of $\Pi\otimes\WW_{-5/2}(\cf_2, f_{\rm prin})$.

Remarkably, this construction can then be generalised to a realisation of the general subregular algebra $\WW^k(\cf_2, f_{\rm subreg})\subset\Pi\otimes \WW^k(\cf_2, f_{\rm prin})$. Taking $W_2$ and $W_4$ to generate a (generic-level) $\WW^k(\cf_2, f_{\rm prin})$ with $W_2$ a Virasoro operator of central charge
\begin{equation}
c_{\WW[2,4]}=-\frac{(3\kZ+2)(5\kZ+8)}{4+\kZ}~,
\end{equation}
we have
\begin{widetext}
\begin{equation}
\begin{split}
X &= e^{\delta+\varphi}~,\qquad Z = \tfrac{\kZ}{2}\partial\varphi~,\qquad T = W_2 + T_{\varphi} + T_{\delta}~,\\[4pt]
Y &= \left(
(\partial\delta)^4
-\tfrac{8(4+\kZ)}{\kZ^2}W_2(\partial\delta)^2
+\tfrac{16(8+6\kZ+\kZ^2)}{\kZ^3}W_2\partial\delta
-\tfrac{8(1+\kZ)}{\kZ}\partial^2\delta(\partial\delta)^2
+\tfrac{16+20\kZ+7\kZ^2}{\kZ^2}(\partial^2\delta)^2
\right.\\
&\left.\qquad\qquad\qquad\qquad\qquad\qquad\qquad\qquad
+\tfrac{8(4+\kZ)}{\kZ^2}\partial W_2\partial\delta+\tfrac{8+6\kZ}{\kZ}\partial^3\delta\partial\delta
-\tfrac{2(4+7\kZ+3\kZ^2)}{3\kZ^2}\partial^4\delta+\tfrac{4+\kZ}{\kZ^4}\OO^{(4)}_{W}
\right)e^{-\delta-\varphi}~,\\[4pt]
\OO^{(4)}_W &= \tfrac{32 (4 + \kZ) (40 + 79 \kZ + 24 \kZ^2)}{75\kZ^2+148\kZ-8}W_2^2-\tfrac{16 (96 + 212 \kZ + 133 \kZ^2 + 59 \kZ^3 + 15 \kZ^4)}{75 \kZ^2+148\kZ-8}\partial^2 W_2 +
\tfrac{1440 (-128 - 112 \kZ + 24 \kZ^2 + 26 \kZ^3 + 3 \kZ^4)}{75 \kZ^2+148\kZ-8}W_4~.
\end{split}
\end{equation}
\end{widetext}
In this equation, $\kZ$ denotes the level of the $U(1)$ current in $\WW^k(\cf_2, f_{\rm subreg})$ which is related to the level of the $\cf_2$ affine current algebra by $\kZ=4(k+2)$ and to the normalization of $\varphi$ by $\langle \varphi,\varphi\rangle=4/\ell$.

When specialised to the value $\kZ=-\tfrac{8}{5}$, the entire $\WW^k(\cf_2, f_{\rm prin})$ algebra becomes null and its simple quotient $\WW_k(\cf_2, f_{\rm prin})$ becomes trivial. Consequently we recover a free field realisation of $\WW_{-12/5}(\cf_2, f_{\rm subreg})$ using only the lattice degrees of freedom from $\Pi$ (in fact, only the subalgebra with even $\rmU(1)_F$ charge). This then gives a free field realisation of precisely the VOA associated to the $(A_1,A_7)$ Argyres-Douglas SCFT \cite{Beem:2017ooy}, for which the Higgs branch is still $\bC_2/\bZ_4$ but there are no additional residual degrees of freedom in generic Higgs vacua.

It is natural to ask if there is a more general story of free field realisations for Drinfel'd--Sokolov $\WW$-algebras along these lines, and indeed this example seems to be a particular instance of a more general (conjectural) construction \cite{BMLprogress}. We summarise the apparent situation as follows: given a finite-dimensional, semisimple Lie algebra $\gf$ and two nilpotent elements (representing conjugacy classes) $f$ and $f'$, such that $f'$ covers $f$ \footnote{Conjugacy classes of nilpotent elements of $\gf$ form a partially ordered set. Here we adopt standard terminology in saying that $y$ \emph{covers} $x$ if $y \succ x$ and there is no $z$ such that  $y \succ z \succ x$, with $x$, $y$ , and $z$ elements of the poset.}, then (we conjecture) there exists a realisation (sometimes referred to as a generalised free field realisation) of $\WW^k(\gf,f)$ as a subVOA of $\bV_{\text{free}}\otimes \WW^k(\gf, f')$ where the factor $\bV_{\text{free}}$ is a lattice VOA. These free fields have a transparent geometric interpretation as they are associated to the transverse space to the nilpotent orbit $\mathbb{O}_{f}$ in $\overbar{\mathbb{O}}_{f'}$. 

In the example above, we have precisely this structure where $\gf=\cf_2$, $(f,f')=( f_{\rm subreg}, f_{\rm prin})$, and the transverse space in question is $\mathbb{C}^2/\mathbb{Z}_4$, see \cite{Kraft1982OnTG}. The case of $\gf=\af_2$, $(f,f')=( f_{\rm subreg}, f_{\rm prin})$ appeared recently in \cite{Adamovic:2020ktz} where a realisation of the Bershadsky--Polyakov algebra ${\rm BP}^k\cong\WW^k(\af_2, f_{\rm subreg}= f_{\rm min})$ was given. That same construction, together with one corresponding to the pair $(f,f')=(0,f_{\rm min})$, was produced independently in \cite{BMLun}. For arbitrary $\gf$ and $(f,f')=(0,f_{\rm min})$ the realisation is obtained by a simple generalisation of the one presented in \cite{Beem:2019tfp}. For the most general choices of $\gf$ and $f,f'$, some technical obstacles and conceptual challenges remain. We will return to this briefly in Section \ref{sec:discussion}.

\vspace{-10pt}

\section{\label{sec:Higgs_branch_model}Recovering Higgs and Hall-Littlewood chiral rings}

\vspace{-2pt}

A simple application of the free field realisation given above is to derive/predict the structure of the Hall-Littlewood chiral ring by taking a semi-classical limit. As in \cite{Beem:2019tfp}, the free field vertex algebra used here admits a canonical, ascending $R$-filtration that we propose to identify with the $R$-filtration coming from the underlying $\rmSU(2)_R$ symmetry of the four-dimensional SCFT (\cff\ \cite{Beem:2017ooy}). For the isotropic lattice algebra $\Pi$, the filtration is defined by (here $j_\pm\colonequals \frac12(\partial\delta\pm\partial\varphi)$),
\begin{equation}
\begin{split}
F_R\Pi &={\rm span}\left\{(j_{\pm})^{n^{\pm}_0}(\partial j_{\pm})^{n^{\pm}_1}\cdots(\partial^{k}j_{\pm})^{n^{\pm}_k}e^{\frac{m}{2}(\delta+\varphi)} \right\}~,\\
&~~~~{\rm where}~n^-_0+\ldots+n^-_k + m \leqslant R~.
\end{split}
\end{equation}
For the symplectic fermions we have,
\begin{equation}
\begin{split}
F_R\mathrm{SF}(2) &={\rm span}\left\{\prod_{A=1}^4(\eta_{A})^{n^{A}_0}\cdots(\partial^{k}\eta_{A})^{n^{A}_{k_A}} \right\}~,\\
&~~~~{\rm where}~\sum_{A=1}^{4}n^A_0+\ldots+n^A_{k_A} \leqslant \frac12 R~.
\end{split}
\end{equation}
For both of these, the normally ordered product descends to a graded (super)commutative multiplication and the singular terms in the OPE reduce to a vertex Poisson algebra structure at the level of the associated graded. By further restricting to the subspace of the associated graded with $h=R$ we recover the Higgs branch chiral ring as a Poisson algebra, with the Poisson bracket encoded in simple poles. In the case at hand, this amounts to replacing $j_-\to\mathsf{2p}$, $j_+\to0$, $e^{\frac{n}{2}(\delta+\varphi)}\to\mathsf{e}^{\frac{n}{2}}$ and dropping all terms with additional differentiation as well as the entire symplectic fermion algebra. Additionally, $\mathsf{p}$ and $\mathsf{e}$ inherit a canonical Poisson structure $\{\msf{p},\msf{e}\}=\msf{e}$.

Applying these replacements to the VOA generators, we set to zero all but $\{X, Y, Z\}$, which correspond to $\hat{\BB}$ operators, and for those the replacements return a ``free-field realisation'' of the Higgs branch chiral ring,
\begin{equation}
\begin{split}
X &~~\to~~ {\mathsf x} = \mathsf{e}~,\\
Z &~~\to~~ {\mathsf z} = \mathsf{2\,p}~,\\
Y &~~\to~~ {\mathsf y} = \mathsf{16\,p^4e^{-1}}~,
\end{split}
\end{equation}
These manifestly reproduce the defining Higgs chiral ring relation and give us back the expressions in terms of the coordinate ring on $\UU_{\msf{x}}$ from \ref{subsec:higgs_branch}.

In the presence of operators with nonvanishing $\rmU(1)_r$ charge, we can instead make a more refined restriction of the associated graded to components obeying the relation $h=R+r$. Doing this returns a kind of free-field realisation of the Hall-Littlewood chiral ring. (Alternatively, one could restrict to components with $h=R-r$ and recover the Hall-Littlewood antichiral ring, which is isomorphic.) At the level of our free field generators, this amounts to addititionally retaining (undifferentiated) $\eta_{1}$ and $\eta_2$ (with trivial Poisson bracket), while setting to zero $\eta_3$ and $\eta_4$ as well as all derivatives of symplectic fermions. Of course, this breaks the $\rmUSp(4)$ outer automorphism symmetry to $\rmU(2)$. We are left with the following additional $\DD$-type generators of the Hall-Littlewood chiral ring
\begin{alignat}{2}
b^+_{~\;a} = \,\chi^+_a\, ~\to~~ & \upchi^+_a&~=~&\upeta_a\mathsf{e}^{\frac{1}{2}}~,\nonumber\\
b^-_{~\;a} = \,\chi^-_a\, ~\to~~ & \upchi^-_a&~=~&4\,\upeta_a\mathsf{p}^2\mathsf{e}^{-\frac{1}{2}}~,\\
\omega = \Upsilon_{12} ~\to~~ &\Upupsilon&~=~&\upeta_1\upeta_2\mathsf{p}~,\nonumber
\end{alignat}
where $a=1,2$. From these expressions, one can easily read off chiral ring relations amongst these generators, as well as a Poisson bracket for the Hall-Littlewood chiral ring that extends that of the Higgs branch chiral ring. Indeed, we have confirmed that these reproduce the chiral ring relations and Poisson brackets as they are computed directly from the Lagrangian description \cite{CJBDBS:HL}.

\vspace{-10pt}

\section{\label{sec:discussion}Discussion}

\vspace{-2pt}

The realisation of the genus-two VOA presented here has a number of suggestive qualities that warrant further comment.

\medskip

The most prominent feature of the construction is the manifestation of the large outer automorphism group $\rmUSp(4)$. Even more so than the $\rmSU(2)$ subgroup that was identified in \cite{Kiyoshige:2020uqz}, this is an unlikely looking symmetry since it mixes four-dimensional multiplets of much different types ($\DD$,$\overbar{\DD}$, and $\hat\CC$). Such a phenomenon has been seen to arise in a few previously studied examples, namely in the free vector multiplet theory (corresponding to just symplectic fermions with the attendant symplectic outer automorphism group), and in $\mathcal{N}=4$ supersymmetric Yang-Mills theory, where there is a $\rmUSp(2)$ outer automorphism symmetry of the small $\mathcal{N}=4$ superconformal algebra that seems to be preserved in the full VOA \cite{Beem:2013sza,Bonetti:2018fqz,Beem:2019tfp}. Based on these examples, it seems the existence of such outer automorphisms may be meaningfully related to the existence of an enhanced Higgs branch.

It is further interesting to observe that this symmetry group appears as the flavour symmetry of the \emph{three-dimensional} Coulomb branch obtained by reducing our genus-two SCFT on a circle and flowing to the infrared SCFT. That three-dimensional fixed point has a mirror dual description as an $\rmSU(2)$ gauge theory with two hypermultiplets both transforming in the adjoint representation of the gauge group \cite{Benini:2010uu}, a description that renders the $\rmUSp(4)$ symmetry transparent. A (potentially trivial) action of Coulomb branch symmetries by outer automorphisms arises for VOAs realised on boundaries of three-dimensional $\NN=4$ SCFTs \cite{Gaiotto:2016wcv,Costello:2018fnz}, and the present theory can certainly be realised as such a boundary VOA by compactification on a cigar (or alternatively directly in terms of the three dimensional reduction in the $H$-twist). However, the $\rmUSp(4)$ action we have found seems difficult to interpret in this manner, since it includes as a subgroup the $\rmU(1)_r$ symmetry. It would of interest to better understand the interpretation of this outer automorphism from complementary perspectives.

\medskip

It has been proposed in \cite{Beem:2019tfp,Beem:2019snk} that free field realisations of this type should be thought of in four-dimensional terms as a kind of inverse Higgsing operation, and the present example seems to reinforce this idea. In the cases studied in those previous works, the Higgsing operation in question could be understood at the level of VOAs as an instance of quantum Drinfel'd Sokolov reduction for an $\widehat{\slf(2)}$ affine subalgebra of the parent VOA. The free field realisations were then performing an inverse Drinfel'd--Sokolov reduction. 

In the present case, we would interpret the Higgsing operation that is being inverted amounts as giving ${\sf x}$ an expectation value (which preserves a ${\rm diag}(\rmU(1)_F\times\rmU(1)_R)$ symmetry that is interpreted as the infrared $\rmU(1)_R$ symmetry) leading to the restriction to the patch $\UU_{\sf x}\subset\MM_{\rm H}$). It is tempting to conjecture that this Higgsing can be implemented at the level of the VOA by introducing a $(0,1)$ $bc$-ghost system and passing to the cohomology of a BRST operator of the form
\begin{equation}
Q_{\rm BRST} = \int\frac{dz}{2\pi i}c(z)(X(z)-1)~.
\end{equation}
One may quickly check that a modified version of the stress tensor,
\begin{equation}
T_{\rm IR} = T + T_{bc} +\partial Z~,
\end{equation}
is $Q_{\rm BRST}$ closed and has $c=-4$, but we have not attempted a more detailed analysis. The full BRST cohomology should be identifiable with the symplectic fermion VOA. We are not aware of any analysis of precisely this type of BRST problem in the VOA literature.

\medskip

As we mentioned in Section \ref{sec:observations}, our realisation of the generic-level $\WW^k(\cf_2,f_{\rm subreg})$ is in many ways analogous to the realisation of the Bershadsky-Polyakov algebra $\WW^k(\af_2,f_{\rm subreg})$ presented in \cite{Adamovic:2020ktz}. In that work, the realisation was used to understand the structure of a fairly general class of representations of the VOA known as relaxed highest weight modules. It would be interesting to study the same class of modules in the present case, and potentially connect with the study of class $\SS$ surface operators.

\medskip

An important problem is to generalise the free-field methods described in this and previous papers of the authors to the point where they accommodate the higher-genus class $\SS$ theories. At genus $g$, the Higgs branch is the Kleinian singularity of type $D_{g+1}$, so in particular for $g\geqslant3$ there are no flavour symmetries. The absence of an organising $\rmU(1)$ symmetry is an apparent obstacle to implementing the same sort of free field scheme used in this paper. (Indeed, the difficulties associated with the general $D$-type Kleinian singularities arise in the context of the conjectural Drinfel'd--Sokolov programme outlined at the end of Section \ref{sec:observations}.) However, for the higher-genus theories it is reasonable to suspect that other features of the present example may generalise more easily. 

A natural conjecture is that the higher-genus VOAs will enjoy $\rmUSp(2g)$ outer automorphism symmetries and that the $\rmUSp(2g)$ invariant subVOA will be isomorphic to $\WW_{-g-1/2}(\cf_g, f_{\rm subreg})$ and realised in terms of ${\rm SF}(2g)^{\rmUSp(2g)}\simeq \WW_{-g-1/2}(\cf_g, f_{\rm reg})$ and a rank $(1,1)$ lattice VOA. An analysis of the superconformal index along with results of \cite{CJBDBS:HL} suggest that the these theories have $\DD$ and $\overbar{\DD}$-type generators of dimension $h=2$ and $h=g$ that transform in the fundamental $(2g)$-dimensional representation of the outer automorphism group (generalising $\chi^\pm$ at genus two). Hopefully these will (weakly) generate the full VOA, allowing for a similarly economical derivation of the full VOAs to what we found here.

\medskip

\noindent We hope to return to all of these points in future work.

\begin{acknowledgments}
The authors would like to thank Diego Berdeja Su\'arez, Thomas Creutzig, Sujay Nair, and Wolfger Peelaers for useful discussions and collaborations on this and related topics. We would especially like to thank Leonardo Rastelli for collaboration on the larger project concerning free field realisations of four-dimensional VOAs. The research of C.B. is supported in part by ERC Consolidator Grant \#864828 ``Algebraic Foundations of Supersymmetric Quantum Field Theory (SCFTAlg)'' and in part by the Simons Collaboration for the Nonperturbative Bootstrap under grant \#494786 from the Simons Foundation.  The work of C.M. has is funded by the European Union’s Horizon 2020 research and innovation programme under the Marie Skłodowska-Curie grant agreement \#754496.
\end{acknowledgments}
\bibliography{genus2FFR}
\end{document}